\documentclass[amsmath,superscriptaddress,reprint]{revtex4-1}

\usepackage{hyperref}
\usepackage{graphicx}

\begin{document}

\title
{Localized guided-mode and cavity-mode double resonance in photonic crystal nanocavities}
\author{X.~Liu}
\author{T.~Shimada}
\author{R.~Miura}
\affiliation{Institute of Engineering Innovation, 
The University of Tokyo, Tokyo 113-8656, Japan}
\author{S.~Iwamoto}
\author{Y.~Arakawa}
\affiliation{Institute of Industrial Science, 
The University of Tokyo, Tokyo 153-8505, Japan}
\author{Y.~K.~Kato}
\email[Corresponding author: ]{ykato@sogo.t.u-tokyo.ac.jp}
\affiliation{Institute of Engineering Innovation, 
The University of Tokyo, Tokyo 113-8656, Japan}

\begin{abstract}
We investigate the use of guided modes bound to defects in photonic crystals for achieving double resonances.  Photoluminescence enhancement by more than three orders of magnitude has been observed when the excitation and emission wavelengths are simultaneously in resonance with the localized guided mode and cavity mode, respectively. We find that the localized guided modes are relatively insensitive to the size of the defect for one of the polarizations, allowing for flexible control over the wavelength combinations. This double resonance technique is expected to enable enhancement of photoluminescence and nonlinear wavelength conversion efficiencies in a wide variety of systems.
\end{abstract}

\maketitle

Photonic crystals allow high degree of control over light through periodic modulation of refractive index \cite{Joannopoulos}. In particular, the photonic band gap, which inhibits propagation of light with a frequency within the gap, can be used to confine light and form optical nanocavities with state-of-the-art quality factors exceeding one million \cite{Takahashi:2007, Sekoguchi:2014}. Combined with their small mode volumes, these nanocavities are ideal for coupling to a variety of nanoscale emitters \cite{Yoshie:2001, Fushman:2005, Englund:2010, Watahiki:2012}. By matching the luminescence wavelength to the cavity resonance, emission rates can be enhanced and radiation patterns can be redirected to achieve higher efficiencies \cite{Noda:2007, Fujita:2008}.

When the emitters are optically excited, further control can be achieved by tuning another mode in a cavity to the excitation wavelength to obtain a double resonance \cite{Ota:2008, Imamura:2013}. Such a double resonance would also be desirable for nonlinear wavelength conversion processes such as sum and difference frequency generation, four-wave mixing, and Raman scattering. It is, however, a challenge to match two cavity modes to a specific pair of wavelengths, as they are usually not independently tunable. A recent demonstration of a silicon Raman laser takes advantage of modes with different symmetry to fine-tune the double resonance \cite{Takahashi:2013}, while polarization can also be used to control the mode separation \cite{Zhang:2009, McCutcheon:2011}. For independent tuning of resonances at large wavelength differences, a cross-beam design has also been used \cite{Rivoire:2011apl, Rivoire:2011oe}.

Here we characterize guided-mode resonances localized at defects in photonic crystals using photoluminescence (PL) microscopy for their use in achieving double resonances. Unlike the cavity-mode resonances, these localized guided modes are at frequencies outside the photonic band gap, and large wavelength separation from the cavity modes is possible. When both the excitation and the emission wavelengths coincide with the localized guided mode and the cavity mode, respectively, we observe PL enhancement factor as high as 2400. Changing the defect structure does not cause significant shifts in the localized resonances for one of the polarizations, allowing for a simple design procedure to achieve double resonances for various wavelength combinations. We demonstrate such flexibility by tuning the double resonance in a wide range of excitation and emission wavelengths.

\begin{figure}
\includegraphics{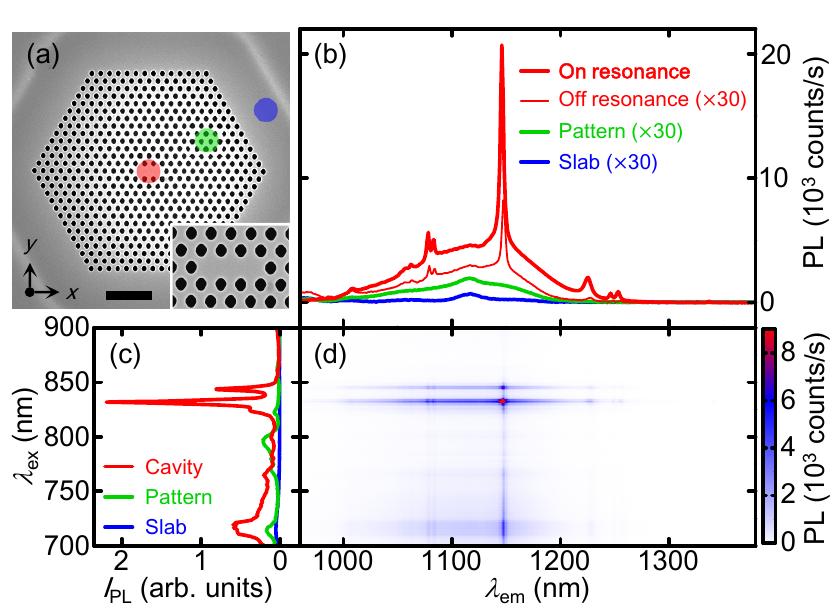}
\caption{\label{fig1}
(a) A scanning electron micrograph of a device with an L3 cavity and $a=370$~nm. The scale bar is 2~$\mu$m. Red, green, and blue dots indicate the positions at which the curves with the corresponding colors in (b) and (c) are taken. Inset shows a magnified view of the cavity.
(b)Red, green, and blue thick solid lines show the PL spectra taken at the cavity, patterned area, and unetched slab area, respectively, with $\lambda_\text{ex}$ = 832~nm. The thin red line shows a PL spectra taken at the cavity with $\lambda_\text{ex} = 900$~nm. Green, blue, and thin red curves have been multiplied by 30 for clarity. 
(c) PLE spectra taken at the cavity (red), patterned area (green), and unetched slab area (blue). PL intensities in (d) have been integrated between $\lambda_\text{em} = 1090$~nm and 1130~nm to obtain these PLE spectra.
(d) A PLE map taken at the cavity. All data in (b-d) are taken with $x$-polarized excitation with $P=1$~mW.
}\end{figure}

Our devices are L-type cavities in hexagonal lattice photonic crystal slabs made from silicon-on-insulator wafers \cite{Iwamoto:2007, Watahiki:2012}. Electron beam lithography followed by a dry etching process defines the air holes in the 200-nm-thick top Si layer. We have designed the air hole radius to be $r=0.29 a$ where $a$ is the lattice constant. Linear defects are introduced to form L$n$ cavities, where $n$ denotes the number of missing holes. Subsequent selective wet etching removes the 1000-nm-thick buried oxide layer, thereby forming a free-standing photonic crystal membrane. Scanning electron micrographs of a typical device are shown in Fig.~\ref{fig1}(a).

The cavities are characterized with a laser-scanning confocal microscope \cite{Moritsubo:2010, Watahiki:2012}. A wavelength-tunable continuous-wave Ti:sapphire laser allows for PL excitation (PLE) spectroscopy, and PL images are acquired by scanning the laser beam with a fast steering mirror. An objective lens with a numerical aperture (NA) of 0.8 focuses the excitation laser beam on the sample to a spot size of $\sim$1~$\mu$m, and the same lens collects the PL from silicon. Linear polarization of the excitation laser can be rotated using a half-wave plate placed just before the objective lens. A single-grating spectrometer disperses PL onto a liquid-nitrogen-cooled InGaAs photodiode array for detection. All measurements are done in air at room temperature.

Typical PL spectra are shown in Fig.~\ref{fig1}(b) for excitation with its electric field $E$ polarized along the $x$-axis and an incident power $P=1$~mW. The thin red line is the spectrum taken with the laser spot on the cavity at an excitation wavelength $\lambda_\text{ex} = 900$~nm, clearly showing a cavity mode at an emission wavelength $\lambda_\text{em} = 1148$~nm with a quality factor $Q=360$, which is assigned to the 5th mode of the L3 defect \cite{Fujita:2008}. Interestingly, when $\lambda_\text{ex}$ is tuned to 832~nm, we observe $\sim$50-fold increase in the PL intensity throughout the spectrum (thick red line). 

In order to study this effect in detail, PLE spectroscopy is performed [Figs.~\ref{fig1}(c) and \ref{fig1}(d)]. We see a clear ``cross'' pattern in the PLE map taken at the cavity [Fig.~\ref{fig1}(d)], where the vertical line corresponds to the cavity mode in resonance with emission, while the horizontal line represents a resonance in the excitation wavelength. At the intersection, we have a double resonance, where both excitation and emission are resonant. In Fig.~\ref{fig1}(b), the double resonance condition is met at the strongest peak in the thick red curve, and we find that the detected PL peak intensity is enhanced by a factor of $\sim$2400 compared to an unetched part of the Si slab (blue curve) at the same wavelength. 

We characterize the excitation resonance using PLE spectra obtained by plotting the integrated PL intensity $I_\text{PL}$ as a function of $\lambda_\text{ex}$ [Fig.~\ref{fig1}(c), red curve]. The spectral integration has been performed over a 40-nm window centered at $\lambda_\text{em} = 1110$~nm to eliminate the effects of the cavity modes. There are several sharp peaks on top of a broad spectral feature, and the strongest excitation resonance at $\lambda_\text{ex} = 832$~nm has $Q = 290$. 

Such high-$Q$ resonances cannot be attributed to cavity modes that are confined by the photonic band gap, as the high-energy band edge is located at around 1100~nm for this lattice constant \cite{Fujita:2008}. At shorter wavelengths, guided modes exist which allow photons to freely propagate in the plane of the photonic crystal slab. These guided modes can couple to free-space modes if they exist above the light line \cite{Ochiai:2001} and can enhance light emission over a large area because they are delocalized \cite{Fujita:2008, Noda:2010}.

We evaluate the effects of the guided modes by performing measurements within the photonic crystal pattern but away from the cavity. The green curve in Fig.~\ref{fig1}(b) shows a PL spectrum with some enhancement compared to the emission from the unetched slab area, although the enhancement is significantly smaller than that observed at the cavity. A PLE spectrum on the pattern [Fig.~\ref{fig1}(c), green curve] shows a broad spectral feature which is expected for guided modes because they have angle-dependent frequency \cite{Ochiai:2001} and excitation with a high NA lens would couple to a wide range of frequencies. We note that the high-$Q$ resonances are absent in this spectrum, suggesting that they are only observed near the defect.

\begin{figure}
\includegraphics{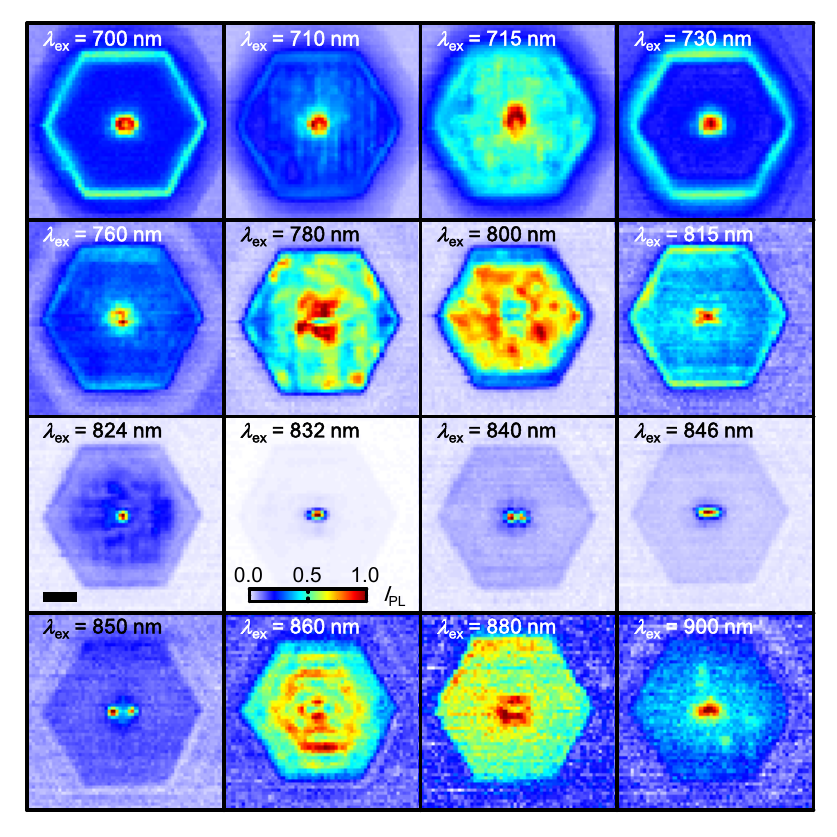}
\caption{\label{fig2}
PL images of the device shown in Fig.~\ref{fig1} at various $\lambda_\text{ex}$, taken using $x$-polarized excitation with $P=1$~mW. The images are constructed by integrating the PL intensity within a 40~nm window centered at 1110~nm and normalizing by the maximum value. The scale bar is 2 $\mu$m. All panels share the scale bar and the color scale.
}\end{figure}

The spatial extent of the high-$Q$ excitation resonances are investigated by performing imaging measurements. In Fig.~\ref{fig2}, we present PL images taken at 16 different excitation wavelengths. As the emission spectral integration window has been chosen not to include the cavity modes, these images mostly reflect the excitation efficiency profiles. The images change drastically for the different wavelengths, and an image taken at $\lambda_\text{ex} = 800$~nm shows that the PL enhancement occurs throughout the photonic crystal pattern, consistent with the delocalized nature of the guided modes.  In comparison, it is clear that the high-$Q$ excitation resonances at $\lambda_\text{ex} = 832$~nm and 846~nm are highly localized at the defect. Similar localized resonances can be seen at other wavelengths as well, although the PL enhancements are smaller.

\begin{figure}
\includegraphics{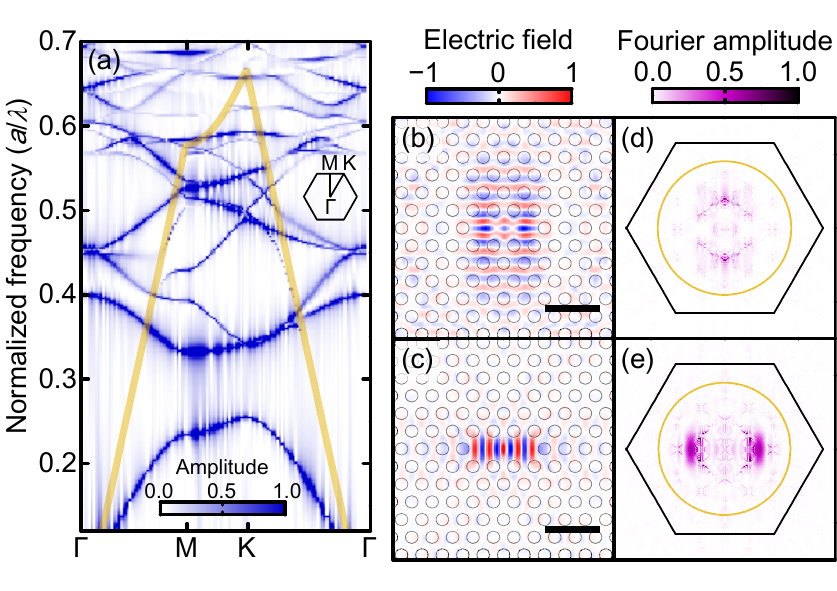}
\caption{\label{fig3}
(a) Photonic band diagram of transverse-electric modes computed for a defect-free structure with $r/a =0.305$ and a slab thickness of $0.54 a$. Temporal Fourier transforms of impulse response are plotted, where wave vectors are specified by appropriate boundary conditions. Refractive index $n= 3.6$ is used for Si, and a slightly larger value of $r/a$ as determined from electron micrographs is used. The orange line indicates the light line.
(b) and (c) Simulated electric field amplitude patterns of localized guided-modes with $x$- and $y$-polarization, respectively. Scale bars are 1~$\mu$m and both panels share the color scale at the top. $E_x$ and $E_y$ are plotted for the $x$- and $y$-polarized modes, respectively. The $x$-polarized mode is at $a/\lambda=0.454$ with $V=4.92 (\lambda/n)^3$ and $Q=300$, while the $y$-polarized mode is at $a/\lambda=0.449$ with $V=2.05 (\lambda/n)^3$ and $Q=100$. 
(d) and (e) Reciprocal space Fourier amplitude maps of the fields shown in (b) and (c), respectively. The orange circle represents the light line, and the hexagon is the first Brillouin zone. 
}\end{figure}

In order to identify the physical origin of the high-$Q$ excitation resonances, three-dimensional finite-difference time-domain (FDTD) calculations are performed.  Figure~\ref{fig3}(a) shows the photonic band diagram for transverse-electric modes in a defect-free hexagonal-lattice photonic crystal slab. The band gap is formed at normalized frequencies $a/\lambda=0.25$ to 0.33, where $\lambda$ is the free-space wavelength. The strongest excitation resonance at $\lambda_\text{ex} = 832$~nm observed in the experiments corresponds to a normalized frequency of 0.445, far above the band gap. Looking at the band diagram near this frequency, several modes exist above the light line around the $\Gamma$ point.  Calculations have shown that these modes can be weakly bound to a defect to form a localized state with a reasonably high $Q$ \cite{Lin:2010}. 

Using FDTD simulations, we have searched for such localized states bound to an L3 defect. Indeed, several modes have been identified, and Fig.~\ref{fig3}(b) shows field distribution for an $x$-polarized resonance with a mode volume $V=4.92 (\lambda/n)^3$ and $Q=300$. The mode profile shows that the field is localized at the defect, consistent with the results of imaging measurements. In addition, the mode has a normalized frequency $a/\lambda=0.454$ which is in the vicinity of the strongest resonance observed in the experiments. Although precise mode assignment is difficult at these frequencies because of dispersion and absorption effects, the characteristics of the excitation resonances can be explained by the simulated mode. With a reasonable agreement between calculations and experiments, we attribute the high-$Q$ excitation resonances to localized guided modes (LGMs). We have also performed calculations for $y$-polarization [Fig.~\ref{fig3}(c)], and have found LGMs at similar frequencies. 

The spatial Fourier transform of the mode profile gives additional insight to the origin of the LGMs. As shown in Fig.~\ref{fig3}(d), the $x$-polarized mode has most of its amplitude within the light line, which implies good coupling to free-space modes. It can also be seen that there are two points with intense amplitudes above and below the zone center. Such a reciprocal space distribution suggests a simple interpretation that this LGM consists of linear combinations of unbound guided-modes. It is reasonable that reciprocal space amplitude shows high intensities at points along the $y$-axis, because these guided-modes are  $x$-polarized. As expected from this picture, the spatial Fourier transform of the $y$-polarized mode [Fig.~\ref{fig3}(e)] shows strong amplitudes for wave vectors along the $x$-axis.

\begin{figure}
\includegraphics{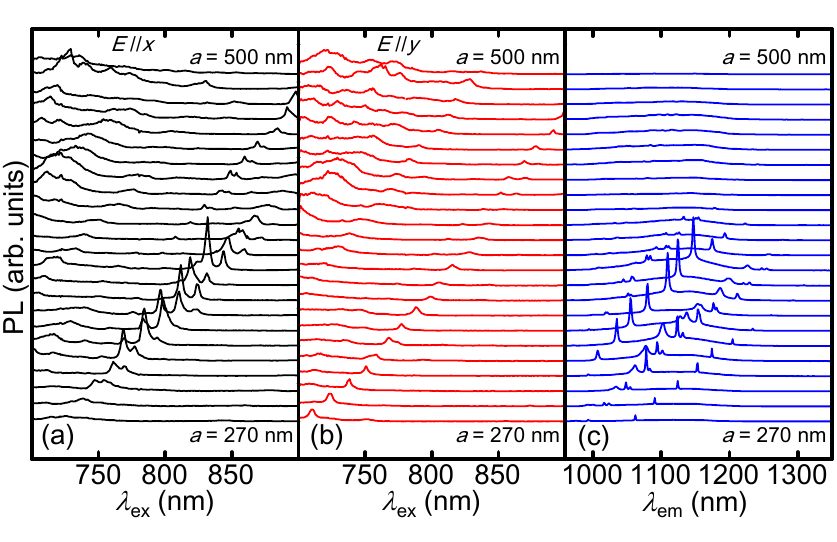}
\caption{\label{fig4}
(a) and (b) PLE spectra for various lattice constants taken under $x$- and $y$-polarized excitation, respectively. (c) Evolution of PL spectra with $a$, taken with $\lambda_\text{ex}$ tuned to the maximum PL intensity and $x$-polarized excitation. The spectra from bottom to top correspond to $a=270$~nm to 500~nm in 10~nm steps. The devices have L3 defects and all spectra are taken with the laser spot on the cavity with $P=1$~mW. 
}\end{figure}

Now we consider how the LGMs can be controlled. We expect the resonances to shift linearly with the lattice constant, as they should depend on the photonic band structure. In Fig.~\ref{fig4}(a), we present $a$-dependence of PLE spectra from L3 defects taken with $x$-polarized excitation. Indeed, the LGMs can be tuned over more than 100~nm by changing the lattice constant. In addition to the two prominent resonances observed for the device with $a=370$~nm, we observe a few other series of weaker resonances. For larger lattice constants, we observe an emergence of a broader spectral structure at shorter wavelengths, likely composed of many peaks. This is reasonable as the band structure becomes more complicated for higher normalized frequencies \cite{Fujita:2008}.

We have also investigated PLE spectra under $y$-polarized excitation [Fig.~\ref{fig4}(b)]. Here we find excitation resonances that scale with $a$ as well, but at slightly different wavelengths compared to those observed for $x$-polarization. For guided modes in a perfect photonic crystal, both polarizations should be degenerate at the $\Gamma$ point where coupling to free-space modes are allowed \cite{Ochiai:2001}. As the defect reduces the symmetry of the system, such an energy splitting for different polarization is expected for the LGMs.

If tuning of a double resonance to a specific combination of two wavelengths is desired, the LGMs and cavity modes need to be controlled independently. When we change the lattice constant, however, the cavity modes also shift linearly [Fig.~\ref{fig4}(c)]. This shows that the resonances cannot be tuned independently, and implies that the double resonance can only be moved along a line in a $\lambda_\text{ex}$-$\lambda_\text{em}$ plane.

\begin{figure}
\includegraphics{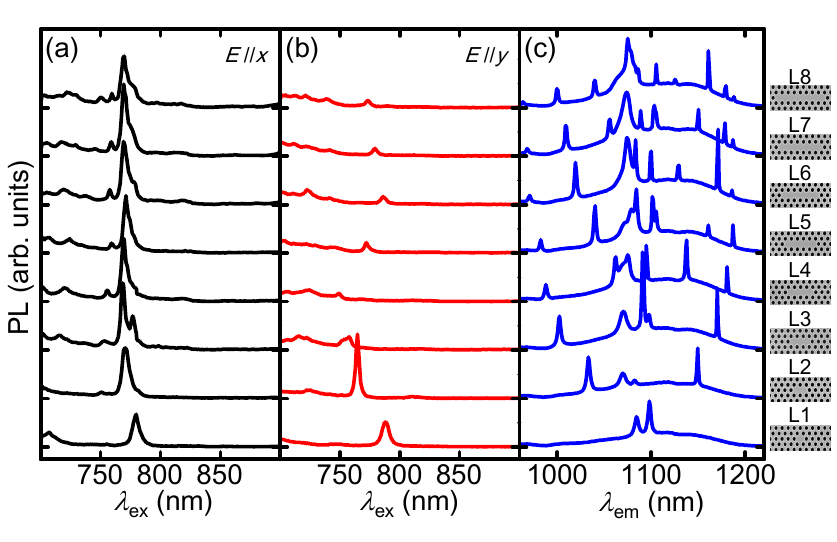}
\caption{\label{fig5}
(a) and (b) PLE spectra of linear defect cavities of various sizes under $x$- and $y$-polarized excitation, respectively. (c) PL spectra of L-type cavities, taken with $\lambda_\text{ex}=710$~nm and $y$-polarized excitation. The spectra from bottom to top correspond to L1 to L8 defects. The devices have $a = 310$~nm and all spectra have been taken at the center of the cavity with $P=1$~mW.
}\end{figure}

In order to find a way to design doubly resonant cavities, we have investigated PLE spectra from linear defects of various lengths. Interestingly, for $x$-polarized excitation, the LGM resonances appear at similar wavelengths despite the different cavity structures [Fig.~\ref{fig5}(a)].  In contrast, the $y$-polarized resonances can differ in wavelength by almost 40~nm [Fig.~\ref{fig5}(b)]. The cavity mode spectra [Fig.~\ref{fig5}(c)] show a complex evolution as the defect length is increased. The number of the modes increases and the resonances appear at different positions within the photonic band gap. 

\begin{figure}
\includegraphics{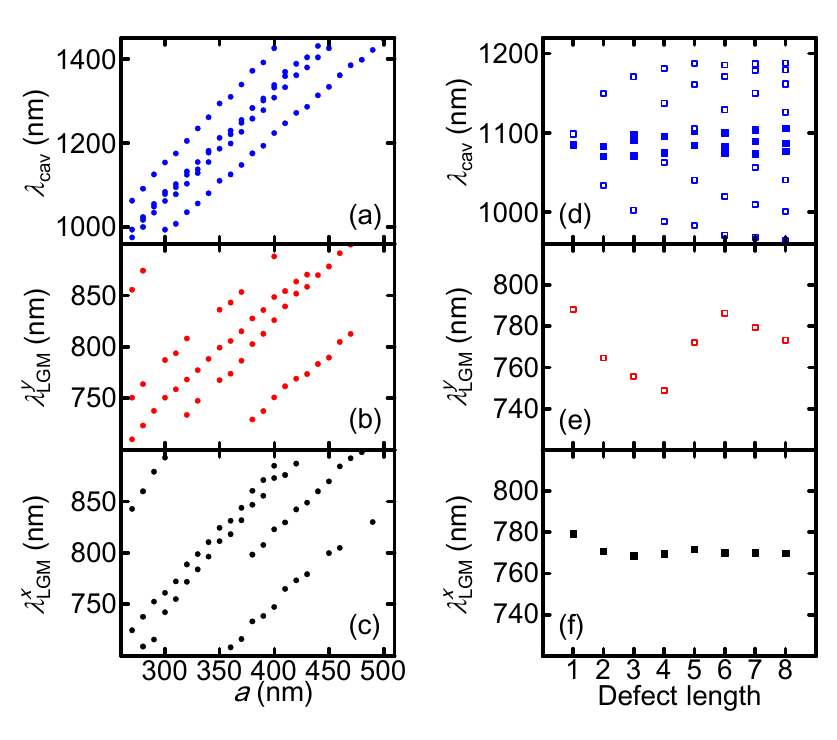}
\caption{\label{fig6}
Lattice constant dependence of (a) cavity modes, (b) $y$-polarized LGMs, and (c) $x$-polarized LGMs for L3 cavities. (d) Defect length dependence of both $x$- (solid symbols) and $y$-polarized (open symbols) cavity modes. (e) and (f) The strongest $y$- and $x$-polarized LGMs, respectively, as a function of the defect length.
}\end{figure}

The lattice constant dependence and the defect length dependence of the LGMs and the cavity modes are summarized in Fig.~\ref{fig6}. We have located the peak positions for $x$-polarized and $y$-polarized LGMs ($\lambda_\text{LGM}^x$ and $\lambda_\text{LGM}^y$, respectively), as well as the cavity resonance wavelength $\lambda_\text{cav}$. Relatively weak intensity peaks that are hard to identify in Fig.~\ref{fig4} are also plotted in Figs.~\ref{fig6}(a-c). The defect length dependence of the cavity modes are shown in Fig.~\ref{fig6}(d), where the mode polarization have been identified by placing a linear polarizer in the collection path. Cavity size dependence of the strongest LGMs are shown in Figs.~\ref{fig6}(e) and \ref{fig6}(f). It is noted that both $x$-polarized LGMs and the cavity modes are relatively insensitive to the defect length, while the $y$-polarized modes show large changes.

Such a different behavior depending on the polarization can be intuitively understood if we consider the LGMs as linear combinations of unbound guided modes. The $x$-polarized modes will consist of guided modes with wave vectors along the $y$-axis, and therefore they should be sensitive to the index of refraction profile along the $y$-direction. Since we are comparing linear defects, the index profiles along the $y$-axis are similar for all cavities, and it is reasonable that the LGMs appear at similar wavelengths. In comparison, for the $y$-polarized modes, the wave vectors will be in the $x$-direction. Changing the length of the defect directly affects the index profile that those guided waves see, and this should result in different resonant wavelengths for the $y$-polarized modes.

The insensitivity of the $x$-polarized LGMs to the length of the defect provides a simple procedure for tuning the double resonance to a specific combination of wavelengths. First, we can choose $a$ to tune an LGM to one wavelength, then we can look for a cavity size which has a mode at the other wavelength. Since the $y$-polarized cavity modes are sensitive to the length of the defect, it should be possible to find a cavity mode near the desired wavelength as long as it is within the photonic band gap. Even though we are changing the defect structure in the latter step, the LGMs would have more or less the same resonance wavelength, allowing for independent tuning of the cavity modes.

\begin{figure}
\includegraphics{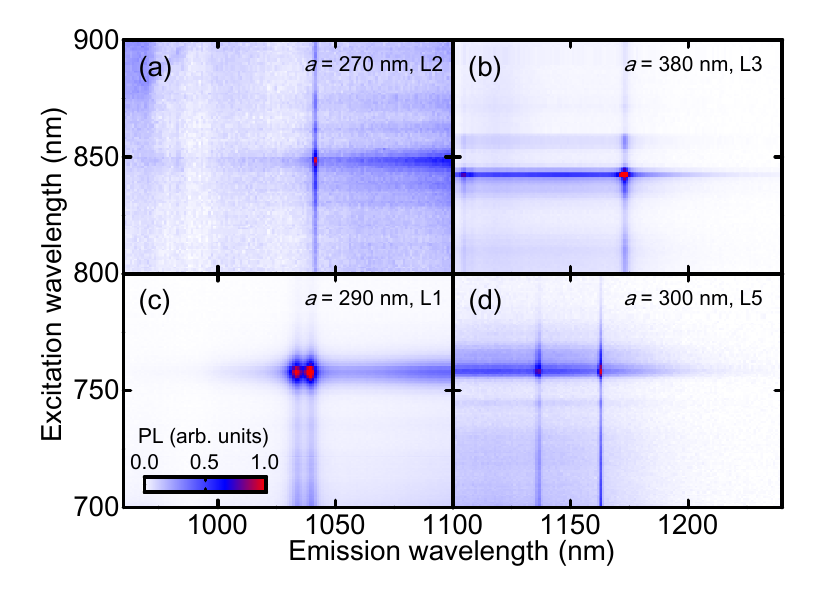}
\caption{\label{fig7}
Double resonance for (a) an L2 cavity with $a=270$~nm, (b) an L3 cavity with $a=380$~nm, (c) an L1 cavity with $a=290$~nm, and (d) an L5 cavity with $a=300$~nm. All data are taken at the center of the cavities using $x$-polarized excitation with $P=1$~mW.
}\end{figure}

We demonstrate such flexibility in Fig.~\ref{fig7} by tuning the double resonance to four different combinations of excitation and emission wavelengths. First, we use the lowest normalized-frequency LGM that we have observed, which is resonant at 850~nm for $a=270$~nm. An L2 defect has a cavity mode at 1040~nm, producing a double resonance in the top-left corner of our PLE maps. Next, we can use an L3 cavity with $a=380$~nm to utilize the strong LGM at 844~nm and the 5th cavity mode at 1173~nm, obtaining a double resonance at the top-right corner. Reducing the lattice constant down to 290~nm, this LGM is now at $758$~nm, and by using an L1 cavity, the double resonance is tuned to the bottom-left corner. Finally, keeping the same LGM for excitation resonance, an L5 defect offers a cavity mode at 1162~nm, completing the last corner. The ratio of the two wavelengths for the double resonance shown in Fig.~\ref{fig7}(d) is 1.53, comparable to those achieved in nanobeam cavities \cite{Rivoire:2011apl, Rivoire:2011oe, Buckley:2014ol, Buckley:2014oe}.

The flexible tuning offered by the LGM and cavity-mode double resonance should allow enhancement of emission from a wide variety of luminescent materials, such as quantum dots, molecules, fluorophores, and proteins. For emitters that show sharp resonances in absorption such as carbon nanotubes \cite{Bachilo:2002}, tuning the double resonance would be particularly effective. It should also be possible to enhance Raman scattering from nanoscale materials such as graphene \cite{Gan:2012}. Increased efficiency in wavelength conversion such as sum and difference frequency generation is expected as well, if materials with large nonlinearity are used \cite{Rivoire:2011oe, Buckley:2014ol}. Although the quality factors of LGMs are smaller compared to typical cavity modes, they may be limited by the strong absorption of silicon. By using a larger electronic band gap material and an enhancement of the $Q$-factors in the L-type cavities \cite{Kuramochi:2014}, further increase in the efficiencies are anticipated.

In summary, we have investigated the use of LGMs for achieving doubly resonant cavities in photonic crystals, and an enhancement of silicon PL by a factor of 2400 has been demonstrated. We have characterized guided modes bound to linear defects in hexagonal lattice photonic crystal slabs using PL spectroscopy and imaging techniques, and found that LGMs have weak dependence on the defect length for one of the polarizations. Taking advantage of such a property, we have shown that the double resonance can be tuned in a flexible manner by choosing a combination of lattice constant and defect length. Our technique offers a simple method for achieving double resonance in a nanocavity, and is expected to be useful for enhancing PL and wavelength conversion at the nanoscale.

\begin{acknowledgments}
Work supported by KAKENHI (24340066, 24654084, 26610080, 26870167), SCOPE, Canon Foundation, Asahi Glass Foundation, and KDDI Foundation, as well as the Project for Developing Innovation Systems, Nanotechnology Platform, and Photon Frontier Network Program of MEXT, Japan. 
\end{acknowledgments}

\bibliography{PCDR}

\end{document}